\documentclass[aps,prl,twocolumn,epsf,showpacs,graphicx]{revtex4}

\usepackage{epsfig,latexsym} 
\usepackage{amsmath,amscd}

\begin{document}

\title{\bf\noindent The statistics of critical points of Gaussian fields
on large-dimensional spaces}
\author{Alan J. Bray$^1$ and David S. Dean$^2$}
\affiliation{$^1$ School of Physics and Astronomy, University of Manchester,
Manchester, M13 9Pl, UK\\ 
$^2$Laboratoire de Physique Th\'eorique (UMR 5152 du CNRS), 
Universit\'e Paul Sabatier, 118, route de Narbonne, 31062
Toulouse Cedex 4, France }
\begin{abstract}
We calculate the average number of critical points of a Gaussian field
on a  high-dimensional space as a  function of their  energy and their
index.  Our results  give a  complete picture  of the  organization of
critical  points  and  are  of  relevance to  glassy and disordered  
systems  and landscape  scenarios  coming from  the  anthropic  
approach to  string theory.
\end{abstract}  
\date{31 October 2006}
\maketitle

The statistics  of random  fields have been  studied in detail  in the
mathematics  and  physics  literature  \cite{adler}.  Gaussian  random
fields are  of particular importance, as  they can be  argued to occur
spontaneously in systems with a  large number of degrees of freedom by
appealing to the central limit theorem. Recently string theorists have
postulated  that the effective  theory arising  from string  theory is
highly complex  leading to  a string landscape  with a huge  number of
possible  minima, each of  which could  describe a  potential universe
\cite{landscape}, thus a statistical study of this complex landscape
may well be necessary. Gaussian  fields 
arise generically  in the study  of classical complex systems  such as
spin glasses,  optimization problems and  protein folding \cite{book}. 
In  many of these systems, jamming  or the onset of a  glass transition 
occurs due
to the presence of a large number of metastable states or local minima
in  the  energy landscape.  High-dimensional  complex potentials  also
occur as  the potential energy  of interacting particle  systems which
exhibit  a  structural  glass  transition  (no  quenched  disorder  is
present).   Many numerical  results  exist on  these potential  energy
landscapes,   notably   for   Lennard-Jones  and soft-sphere 
glass   forming   mixtures \cite{num}.  Scaling  arguments  
relating   the  number  of critical points (points where the gradient of the 
potential energy vanishes)  of  various types also exist 
\cite{indscale}  and the onset of the glass  transition is often identified 
with  the point where the critical  points in  the  free energy  landscape  
become dominated  by minima \cite{saddles}.  In this Letter we compute
the average number of critical points of a Gaussian potential, on a high
dimensional space, at {\em fixed  value of the potential} and 
at {\em fixed number of negative eigenvalues} at the  critical point.

We shall consider a Gaussian field $\phi$ defined  over  volume $V$ 
of an $N$-dimensional Euclidean space. The field has zero mean and
correlation function:       
\begin{equation}
\langle \phi({\bf x}) \phi({\bf y})\rangle = N f\left({({\bf x-y})^2\over 2 N}
\right).
\end{equation}
This scaling in $N$ ensures the existence of the thermodynamic limit
(as $N\to \infty$) of what is known as the toy model, whose statics
\cite{toystat} and dynamics \cite{toydyn} have been studied in the
literature.  The version of the toy model corresponding to our study
here is a single particle in the potential $V({\bf x}) = \phi({\bf
x})$.  In the harmonically confined version of the toy model the
potential energy of the particle is $V({\bf x}) = \phi({\bf x}) +
{1\over 2}\mu {\bf x}^2$, and the presence of the confining potential
means that the volume $V$ of the system can be taken to infinity. The
total number of critical points for the toy model with confining
potential was studied by Fyodorov \cite{fyod} as a function of
$\mu$. Here we study the case without confining potential in a finite
volume. We give explicit expressions for the average of the number,
$\cal{N}(\alpha,\epsilon)$, of critical points as a function of index
$N\alpha$ and their energy $N\epsilon$ in the form $\langle{\cal
N}(\alpha,\epsilon)\rangle \sim \exp[N\Sigma(\alpha,\epsilon)]$,
where $\Sigma(\epsilon,\alpha)$ is the ``complexity'', thereby giving
a complete picture of the geometry of critical points.  Our method can
also be applied to the confined case where $\mu\neq 0$ and
details will be published elsewhere \cite{inprep}.  For the purposes
of this paper we consider the unconfined case as it is more relevant
to landscape scenarios in string theory and other translationally
invariant systems and also because the results are more explicit.

For an  $N\times N$  matrix $H$  the index of  $H$, denoted  by ${\cal
I}(H)$, is defined as the  number of negative eigenvalues of $H$.  The
number of critical  points of energy $E= N\epsilon$  and index $\alpha
N$ is given by the generalized Kac-Rice formula
\begin{eqnarray}
{\cal  N}(\alpha,\epsilon)   &=&  \int_V  d{\bf   x}  \  \prod_{i=1}^N
\delta\left(   \partial_i   \phi({\bf   x})\right)|\det  H({\bf   x})|
\nonumber        \\         &        &\delta\left(\phi({\bf        x})
-N\epsilon\right)\,\delta\left({\cal I} ( H({\bf x}))-N\alpha\right),
\end{eqnarray}
where $H({\bf  x})$ is the Hessian  matrix of the field  $\phi$ at the
point ${\bf x}$, with elements $H_{ij}({\bf x}) = \partial_i\partial_j
\phi({\bf  x})$. The  average of  $\cal{N}(\alpha, \epsilon)$  is then
given, using the spatial translational invariance of the field $\phi$,
as
\begin{eqnarray}
\langle {\cal N}(\alpha,\epsilon)  \rangle &=& V \langle(\prod_{i=1}^N
\delta\left(   \partial_i   \phi({\bf   0})\right))|\det(H({\bf   0})|
\delta\left(\phi({\bf    0})    -N\epsilon\right))\nonumber    \\    &
&\delta\left({\cal I} ( H({\bf 0}))-N\alpha\right)\rangle.
\label{eqav1}
\end{eqnarray}
The non-vanishing correlation  functions for  the Gaussian random  
variables $H({\bf 0})$ and $\phi({\bf 0})$ are:
\begin{eqnarray}
\langle \phi({\bf  0})\ \phi({\bf 0})  \rangle &=& N  f(0) \nonumber\\
\langle  \partial_i\partial_j\phi({\bf 0})\ \partial_k\partial_l\phi({\bf  0}) \rangle
&=& {f''(0)\over N}( \delta_{lk}\delta_{ij} + \delta_{lj}\delta_{ik} +
\delta_{kj}\delta_{il})   \nonumber\\   \langle   \partial_i\partial_j\phi({\bf
0})\ \phi({\bf 0}) \rangle &=& \delta_{ij} f'(0).
\label{corrs}
\end{eqnarray}
Note  that positivity  conditions for  the correlation  function  of a
Gaussian field in Fourier space can  be shown to imply \cite{inprep}  
$f(0) >0$, $f'(0)<0$ and $f''(0)  >0$. When $f$ is an analytic function 
at the origin it is  easy to show that the variables 
$\partial_i\phi({\bf  0})$ are independent of $H({\bf 0})$  and 
$\phi({\bf 0})$, as their correlation with   these  variables   is  
identically   zero.  The   averaging  in
Eq. (\ref{eqav1}) can thus be factorized to give
\begin{eqnarray}  
&&\langle{\cal  N}(\alpha,\epsilon) \rangle  =  V \langle\prod_{i=1}^N
\delta\left(  \partial_i  \phi({\bf 0})\right)\rangle\times  \nonumber
\\&&\langle|\det(H({\bf       0})|      \delta\left(\phi({\bf      0})
-N\epsilon\right))       \delta\left({\cal      I}       (      H({\bf
0}))-N\alpha\right)\rangle.
\label{eqav2}
\end{eqnarray}     
Putting this together yields
\begin{equation}  
\langle  {\cal   N}(\alpha,\epsilon)  \rangle  =   {V\over  \left(2\pi
|f'(0)|\right)^{N\over 2}} \Omega(\alpha,\epsilon),
\end{equation}
where
\begin{equation}
\Omega(\alpha,\epsilon) = \int d\phi \prod_{i\leq j} dH_{ij}\; 
p(H,N\epsilon) |\det(H)| \delta\left({\cal  I}  (H)-N\alpha\right).
\end{equation}
Here  $p(H,\phi)$ is the joint probability density  of $H$ and
$\phi$  (we have   dropped  the  argument  ${\bf
0}$). From the correlators in Eq. (\ref{corrs}) it can be shown that
\begin{equation}
p(H,\phi) = C_N\exp\left(-N^2 S_2(H) - N S_1(H,\phi)\right)
\end{equation}
where $C_N$ is  a normalization constant and the  dominant term of the
{\em action} $S_2(H)$ is given by 
\begin{equation}
S_2(H)  =  {1\over 4  N  f''(0)} \left[  {\rm  Tr} H^2  - {1\over  N}
\left({\rm Tr} H \right)^2\right]
\end{equation}
and the sub-dominant term $S_1(\phi,H)$ is given by
\begin{eqnarray}
S_1(H,\phi)  &=& {1\over  2 f''(0)Q}\left[  {N+2\over N^3}\phi^2  - {2
    f'(0) \over N^2  f''(0)} \phi {\rm Tr}(H)\right. \nonumber  \\ & &
  \left.+ {f(0)\over f''(0) N^2} \left( {\rm Tr}(H)\right)^2 \right],
\end{eqnarray}
where
\begin{equation}
Q= (1 +{2\over N}) {f(0)\over f''(0)} - {f'(0)^2\over f''(0)^2}.
\end{equation}
Positivity  conditions for Gaussian  correlators can  be used  to show
that $Q>0$ for  $N$ finite \cite{inprep}. However it  is possible that
$Q^*=\lim_{N\to\infty}   Q  =   {f(0)\over  f''(0)}   -  {f'(0)^2\over
f''(0)^2}$ is zero. A case,  often used in simulations of diffusion in
random  media \cite{diff},  where this  occurs is  for $f(x)  = \exp(-
x/l^2)$, where $l$ is the correlation length . In this case the action
$S_1$ is  order $N$ and  some modifications of our  approach are
needed. For the moment we will proceed by examining the case $Q^* \neq
0$.  We remark that as  the field $\phi$ is statistically isotropic we
see that the actions $S_2$ and $S_1$ are invariant under the action of
the  rotation  group  $O(N)$  and  that the  associated  measure  thus
describes a Gaussian Orthogonal Ensemble  (GOE).   Notice  that if  we
integrate over the field $\phi$ we find the marginal distribution
for $H$, $p_m(H)$, is given by
\begin{equation}
p_m(H) = C'_N\exp\left(-{N\over  4  f''(0)} 
\left[  {\rm Tr} H^2  -  {1\over N+2}
\left({\rm Tr} H \right)^2\right]\right)
\end{equation}
where $C'_N$ is a normalization constant and
recovers  the result of  Fyodorov \cite{fyod}. Note that  in the
above one can take the limit $Q^* = 0$ without any problems.

We now apply  the standard result for the GOE  where the integral over
$H$ can be reduced to an integral over its eigenvalues \cite{mehta}:
\begin{equation}
p(H,\phi)\prod_{i\leq      j}     dH_{ij}     =      C''_N     
p({\lambda_i},\phi)\prod_{i<j}|\lambda_i -\lambda_j|\prod_{i=1}^N d\lambda_i,
\end{equation}
where $C''_N$  is a  normalization constant.  The  key step is  now to
introduce the normalized density of eigenvalues defined as
\begin{equation}
\rho(\lambda) = {1\over N}\sum_{i=1}^N \delta(\lambda-\lambda_i).
\end{equation}
We now adopt the constrained Coulomb gas approach recently introduced
in \cite{dema} and using standard methods  of functional integration 
\cite{inprep} we may write
\begin{equation}
\Omega(\alpha,\epsilon) = {\Theta(\alpha,\epsilon)\over D}
\label{Omega}
\end{equation}
where
\begin{eqnarray}
&&\Theta(\alpha,\epsilon)  =   \int  d[\rho]  \exp\left(-   N^2  {\cal
S}_2(\rho)  -  N   {\cal  S}_1(\rho,\epsilon)\right)  \nonumber  \\  &
&\delta\left(\int_{-\infty}^0          d\lambda          \rho(\lambda)
-\alpha\right)\delta\left(\int_{-\infty}^\infty          d\lambda          \rho(\lambda)
-1 \right). \label{eqTheta}
\end{eqnarray}
The key point and major  simplification in Eq. (\ref{eqTheta}) is that
the constraint on  the index  becomes  a simple integral
constraint  on the  functional  $\rho$.
The effective actions in Eq. (\ref{eqTheta}) are given by
\begin{eqnarray}
{\cal   S}_2(\rho)   &=&  {1\over   4   f''(0)}\left[  \int   d\lambda
  \lambda^2\rho(\lambda)       -\left(\int       d\lambda      \lambda
  \rho(\lambda)\right)^2\right]\nonumber \\ &-&{1\over 2}\int d\lambda
d\lambda' \rho(\lambda)\rho(\lambda') \ln(|\lambda-\lambda'|)
\end{eqnarray}
and
\begin{eqnarray}
&&{\cal S}_1(\rho,\alpha,\epsilon) = \int d\lambda \rho(\lambda)\left(
\ln\left( \rho(\lambda)\right) - \ln(|\lambda|)\right)\nonumber \\ &+&
{1\over 2  Q f''(0)} \left[  {{N+2}\over N} \epsilon^2  - 2{f'(0)\over
f''(0)} \epsilon \int  d\lambda \lambda \rho(\lambda) \right.\nonumber
\\&+&    \left.{f(0)\over   f''(0)}   \left(\int    d\lambda   \lambda
\rho(\lambda)\right)^2\right].\label{cals1}
\end{eqnarray}
The term $D$ in (\ref{Omega}) is a normalization factor given by
\begin{equation}
D  =  \int  d[\rho]  \exp\left(-   N^2  {\cal  S}_2[\rho]  -  N  {\cal
S}'_1(\rho)\right)
\end{equation}
with
\begin{equation}
{\cal S}'_1(\rho) = {1\over 2 f''(0)}{N\over {N+2}}\left(\int d\lambda
\lambda \rho(\lambda)\right)^2 + \int d\lambda \rho(\lambda)
\ln \rho(\lambda).
\end{equation}
The above functional  integrals can be evaluated by  saddle-point. The
order $N^2$ action  ${\cal S}_2$ has a zero mode  as it is independent
of the average of the eigenvalues,
\begin{equation}
\overline{\lambda} = \int d\lambda \lambda \rho(\lambda).
\end{equation}
Note that  in the usual Gaussian  ensembles the second  term of ${\cal
S}_2$  is absent,  thus if  the average  value of  the  eigenvalues is
shifted from zero  (and thus we shall see later  the value of $\alpha$
is shifted from $1/2$) then the constraint increases the action ${\cal
S}_2$ by a finite value and thus the probability of finding a Gaussian
matrix with $\alpha \neq  1/2$ is of order $\exp(-N^2 \theta(\alpha))$
making the probability of minima for instance extremely small
\cite{indrep,AE,dema}. This is not the case for the Hessian  of 
a Gaussian random field at a critical point due  to the presence of  
this zero mode.  The  degeneracy of the
zero mode is  however lifted by the order $N$  action.  The value
of  $\rho$ minimizing the  action ${\cal  S}_2$ is  given by  a Wigner
semi-circle \cite{wig} law up to an undetermined shift (the zero mode)
\begin{equation}
\rho_{sc}(\lambda) = {1\over 2\pi  f''(0)} \left( 4f''(0) - (\lambda -
{\overline \lambda})^2\right)^{1\over 2},
\label{wigner}
\end{equation}
which  has support  $\lambda\in [{\overline\lambda}  - 2\sqrt{f''(0)},
{\overline\lambda}  +  2\sqrt{f''(0)}]$.  This  distribution  is  {\em
stiff}  and any  departure from  its form  will reduce  the  number of
critical points with index fixed to be non-zero by a factor $\exp(-N^2
\theta(\alpha))$ (with possible $\log(N)$ corrections to the 
prefactor $N^2$ in the case where $\alpha$ is not strictly equal to 
one or zero \cite{indrep}).

Consequently the order $N^2$ terms in the saddle-point expressions for
$\Omega$ and $D$ cancel as  does the order $N$ entropy term $\int
d\lambda   \rho(\lambda)   \ln\left(\rho(\lambda)\right)$,  which   is
clearly   independent  of   $\overline{\lambda}$.   In  addition   the
saddle-point  of $D$  is clearly  at $\overline{\lambda}  =0$. We thus 
find, that  to leading order,  the complexity of critical 
points is given by
\begin{eqnarray}
&&\Sigma(\alpha,\epsilon)={\ln\left({\cal
N}(\alpha,\epsilon)\right)\over   N}\nonumber   \\   &=&  -{1\over   2
f''(0)Q}\left[  \epsilon^2 -  2{f'(0)\over  f''(0)} \overline{\lambda}
\epsilon  +{1\over 2} P{\overline  \lambda}^2\right] \nonumber  \\ &&+
{1\over 2}\ln\left( {f''(0) L^2 \over 2\pi e|f'(0)|}\right),
\end{eqnarray}
where we have made use of the result
\begin{equation}
\int d\lambda \rho_{sc}(\lambda) \ln(|\lambda|) = {\overline \lambda^2
\over 4 f''(0)} + {1\over 2} \left(-1 + \ln(f''(0))\right),
\end{equation}
valid in the regime of interest, $|\bar{\lambda}| \le 2\sqrt{f''(0)}$, and 
have defined
\begin{equation}
P  =  {f'(0)^2\over  f''(0)^2}  + {f(0)\over  f''(0)}(1-  {2\over  N})
\approx {f'(0)^2\over f''(0)^2} + {f(0)\over f''(0)}.
\end{equation} 
In addition, the volume $V$ of the system is written as $V = L^N$, with
$L$ its  characteristic length.  The value of  $\overline{\lambda}$ is
determined from the index constraint
\begin{equation}
\int_{-\infty}^0     d\lambda    \rho_{sc}(\lambda)     =    {2\over
\pi}\int_{{\overline       \lambda       }\over      2\sqrt{f''(0)}}^1
dy(1-y^2)^{1\over 2}  = \alpha.
\label{alplam}
\end{equation}

We now  discuss the consequences of  this general result.   If we only
constrain the energy of the critical points we find a complexity
\begin{equation}
\Sigma(\epsilon)  =   -  {\epsilon^2\over   2  f''(0)  P}   +  {1\over
2}\ln\left( {f''(0) L^2 \over 2\pi e|f'(0)|}\right)
\end{equation}
and the most probable value of $\alpha$ can be computed via
\begin{equation}
{\overline \lambda}(\epsilon) = 2 {f'(0) \epsilon\over f''(0) P},
\label{inprob}
\end{equation}    
and then  using the  relation Eq.\ (\ref{alplam}).   We remark that, as
$f'(0)<0$,  then   as  $\epsilon$  decreases $\overline{\lambda}$
increases   meaning   that    the   most   probable   index   $\alpha$
decreases. Also  for $\overline{\lambda}  > 2 \sqrt{f''(0)}$  the
value  of  $\alpha$  freezes  at  $\alpha =0$,  meaning  that  all  the
eigenvalues are positive and thus  the typical critical points below a
critical energy $\epsilon_c = (f''(0))^{3\over 2} P/f'(0)$ are minima.
If only  the index of the  critical point is constrained  then we find
that
\begin{equation}
\Sigma(\alpha)  = -{{\overline{\lambda}}^2\over  4  f''(0)} +  {1\over
2}\ln\left( {f''(0) L^2 \over 2\pi e|f'(0)|}\right)
\end{equation}
and  that  the most  probable  energy at  a  given  value of  $\alpha$
(equivalently $\overline{\lambda}$) is
\begin{equation}
\epsilon(\overline{\lambda}) = {f'(0)\over f''(0)}{\overline{\lambda}}.
\label{eprob}
\end{equation}
From this we see that the most probable energy of minima, $\epsilon_m$, 
is  $\epsilon_m = 2f'(0)/\sqrt{f''(0)}$. In fact one can show
that $\epsilon_m \geq \epsilon_c$ with equality in the case $Q^*=0$
\cite{inprep}.  This means that generically most minima are at the
energy $\epsilon_m$ but it is only below the energy $\epsilon_c$ that
they are the dominant critical points.  For $\overline{\lambda} > 2
\sqrt{f''(0)}$ all critical points are minima, so the complexity of
minima is 
\begin{equation}
\Sigma_{\rm min} = -1 + {1\over 2} \ln\left( {f''(0) L^2 \over 2\pi
e|f'(0)|}\right).
\end{equation}
Thus there is a characteristic length 
$L_c= \sqrt{{2\pi e^3|f'(0)|/ f''(0)}}$ below which
there are almost surely no minima.     

We now discuss the case where $Q^*=0$. In this case the first line of
Eq. (\ref{cals1}) is of order one but the other terms are of order
$N$. We see, however, that the order $N$ terms only depend on $\rho$
via $\overline{\lambda}$. This means that at the saddle-point $\rho$
still has the Wigner form of Eq (\ref{wigner}). Consequently we find
that
\begin{eqnarray}
\Sigma(\alpha,\epsilon) &=&  -{N\over 4 f(0)}\left[{N+2\over N } 
\epsilon^2 - 2{f'(0)\over f''(0)} \overline{\lambda}
\epsilon +{1\over 2}  P{\overline \lambda}^2\right] \nonumber \\ 
&&+ {1\over 2}\ln\left({f''(0) L^2 \over 2\pi e|f'(0)|}\right),
\end{eqnarray}
where we have   kept the $N$-dependent prefactor of $\epsilon^2$.
Completing the square in $\epsilon$, the above result may be written to 
leading order for large $N$ as
 \begin{eqnarray}
\Sigma(\alpha,\epsilon) &=& - {N\over 4 f(0)}\left(\epsilon-{f'(0)\over f''(0)}
{\overline \lambda}\right)^2 -{{\overline \lambda}^2\over 4 f''(0)} 
\nonumber \\ &+& {1\over 2}\ln\left({f''(0) L^2 \over 2\pi e|f'(0)|}\right),
\end{eqnarray}
which means that when the index is fixed (i.e. $\overline{\lambda}$ is fixed)
then the critical points of index given by $\overline{\lambda}$ are 
concentrated about the energy $\epsilon = \overline{\lambda} f'(0)/f''(0)$ 
as before but the fluctuations about this energy are $O(1/N)$ as opposed to
fluctuations of $O(1/\sqrt{N})$ in the case where $Q^* \neq 0$. 
The same scaling holds for the fluctuations about  the most probable index,
given by Eq. (\ref{inprob}), when the energy is fixed. 

A study of the  p-spin spherical spin glass model  \cite{pspin} has shown
that at energies  below a threshold energy almost  all critical points
are minima.  However, at the  threshold energy all indices  are equally
probable.  In the  model studied  here  there is  a similar  threshold
energy below which  almost all critical points are  minima. However, as
this  energy is  increased the  index  is concentrated  around a  well
defined value up to an upper threshold energy beyond which all critical 
points are maxima.

Numerical investigations of the critical points of glass formers \cite{num}
are carried out by equilibrating the system, freezing it and then 
finding nearby  critical points. In most of these studies a clear linear 
relation is found between the index $\alpha$ and the potential energy
$\epsilon $ for values of $\alpha$ between $0$ and about $0.1$. This is in 
contrast to the results found here where a linear relation exists between 
$\overline\lambda$ and $\epsilon$, and the relation between $\alpha$ and 
$\epsilon$ must be found via Eq. (\ref{alplam}). For small $\alpha$  we
find the relation
\begin{equation}
\alpha \approx {4 \sqrt{2}\over 3\pi}\left| 
{\epsilon -\epsilon_c\over \epsilon_c} \right|^{3\over2}.
\end{equation}
This difference may be due to the fact that the Gaussian landscape is
quite different from that of glass formers. However, it may also be that
the preparation of the system in an equilibrium configuration, before starting 
the critical point search, biases the statistics of the critical points found, 
whereas our calculation is based on a purely flat measure.

In summary, we have provided a complete description of the
distribution of critical points of Gaussian fields in high-dimensional
spaces, as a function of their energy and index.
The generalization to
the case of a confining harmonic potential will be presented elsewhere
\cite{inprep}.  Finally the analytic results 
presented here will  also be useful to  test numerical   methods 
used to explore  the critical points of potentials on large-dimensional spaces.

We thank Satya Majumdar for useful discussions.
AB thanks Universit\'e Paul Sabatier, Toulouse, for its hospitality
while this work was being done.

\end{document}